\documentclass{ptapap}
\usepackage{amssymb}
\usepackage{url}

\author{Katarzyna Małek}[NCBJ,UZG]
\author{Tomasz Krakowski}[NCBJ]
\author{Maciej Bilicki}[LO,UZG,UCT]
\author{Agnieszka Pollo}[NCBJ,UZG,OAUJ]
\author{Magdalena Krupa}[OAUJ,UZG]
\author{Agnieszka Kurcz}[OAUJ,UZG]
\author{Aleksandra Solarz}[NCBJ,UZG]
\affil[NCBJ]{National Centre for Nuclear Research, Otwock, Poland}
\affil[LO]{Leiden Observatory, Universiteit Leiden, The Netherlands}
\affil[UZG]{Janusz Gil Institute of Astronomy, University of Zielona Góra, Poland}
\affil[UCT]{Department of Astronomy, University of Cape Town, South Africa}
\affil[OAUJ]{Astronomical Observatory, Jagiellonian University, Kraków, Poland}
\title{Learning algorithms at the service of WISE survey}

\begin{document}

\maketitle

\begin{abstract}
We have undertaken a dedicated program of automatic source classification in the WISE  database merged with SuperCOSMOS scans, comprehensively identifying galaxies, quasars and stars on most of the unconfused sky. We use the Support Vector Machines classifier for that purpose, trained on SDSS spectroscopic data. The classification has been applied to a photometric dataset based on all-sky WISE 3.4 and 4.6 $\mu$m information cross-matched with SuperCOSMOS B and R bands, which provides a reliable sample of $\sim170$ million sources, including galaxies at $z_{\rm med}\sim0.2$, as well as quasars and stars. The results of our classification method show very high purity and completeness (more than 96\%) of the separated sources, and the resultant catalogs can be used for sophisticated analyses, such as generating all-sky photometric redshifts.

\end{abstract}

\section{Introduction}

Modern wide-field astronomical surveys have cataloged millions of sources, and the forthcoming ones will raise this number to billions. 
The number of object properties available for further analysis, such as photometry, morphology or spectral indexes, also constantly expands, which forces astronomers to work in a multi-dimensional parameter space to classify and study different types of sources.
However, a crucial quantity for reliable identification of various source types -- the spectroscopic redshift -- is usually known only for  a small fraction of the observed objects. For that reason, source classification is most often performed based on photometric properties only.
 
The most basic physical classification separates out foreground stars from extragalactic sources, using morphological parameters: point-like sources are identified as stars, while extended ones as galaxies (e.g.\ \citealt{vasconcellos11}). 
Morphological classification can provide very good results, nevertheless additional information is needed to separate out stars from point-like but extragalactic sources such as quasars or unresolved galaxies. 
The most common, and relatively simple, approach to star-galaxy separation is based on colors.
Galaxies are usually much redder than stars, and using photometric information it may be possible to define separate star and galaxy loci in color-color diagrams (e.g.\ \citealt{pollo10}).
Similar methodology can be applied to find more specific objects such as AGNs, quasars, variable stars or starburst galaxies  \citep[e.g.][]{wozniak04}. 

Nowadays, deep surveys very often focus on detecting high-redshift objects  at faint apparent magnitudes. 
Standard classification methods, employing morphology or colors, are then no longer reliable due to small angular sizes and low signal-to-noise levels.
When dealing with huge source numbers in multi-dimensional parameter space, applying automatized source classification is a natural step for present-day and future astronomy. 

Here we present an application of a machine learning algorithm to identify galaxies, stars and quasars  in a newly compiled dataset, based on two presently the largest all-sky photometric catalogs: the Wide-field Infrared Survey Explorer \citep[WISE;][]{WISE} in the mid-infrared, and the SuperCOSMOS Sky Survey \citep[SCOS;][]{SCOS1} in the optical.
For the classification described here we used the Support Vector Machines (SVM) -- a supervised learning algorithm which is a maximum-margin classifier able to determine decision planes between sets of objects in an $n$-dimensional parameter space. 
This algorithm has already been successfully applied to other datasets \citep{solarz12,saglia12,malek13SVM}. See also Kurcz et al.\ in this volume where SVM-based classification using only WISE information is discussed.

Our goal is to develop automatic object classification in the WISE$\times$SCOS dataset, into three categories: galaxies, quasars, and stars.  
We aim at obtaining (almost) all-sky catalogs of high completeness and purity, based on photometric properties of the sources. This, together with the parallel effort of \cite{Kurcz15} is the first application of the SVM algorithm to a catalog including $\sim 10^8$ sources preselected from the WISE database.

\section{Data and sample selection}
\label{Sec:data}

WISE is a NASA space-based mission, which surveyed the entire sky in four mid-infrared bands, centered at 3.4, 4.6, 12 and 23 $\mu$m ($W1$ -- $W4$ respectively). 
For our analysis we have used the AllWISE dataset\footnote{Available for
download from \url{http://irsa.ipac.caltech.edu/frontpage/}.}, which
combines data from cryogenic and post-cryogenic survey
phases.
It includes $\sim750$ million sources with S/N $\geq$ 5 in at least one band, and its averaged 95\% completeness in unconfused areas is $W1\simeq17.1$, $W2\simeq15.7$, $W3\simeq11.5$ and $W4\simeq7.7$ in Vega magnitudes. 

Our WISE preselection required: 
(1) S/N ratios larger than 2 in the $W1$ and $W2$ bands (the two other channels being too shallow for our purposes); (2) removal of obvious artifacts (\texttt{cc\_flags}[1,2] = DPHO), 
(3) rejection of the Galactic Plane ($|b|<10^\circ$) to avoid confusion; and
(4) $W1< 17$ mag threshold for all-sky uniformity. 
Such a sample selection gives 
343 million WISE sources.

The SuperCOSMOS Sky Survey contains  digitized photographs in three optical bands ($B, R, I$), obtained via  automated scanning of source plates from  the UKST (in the South) and POSS-II (North) observations. 
SCOS measurements of were calibrated using SDSS photometry in the relevant areas, and 2MASS J band over the rest of the sky (Peacock et al., in prep.). 
The publicly available catalog\footnote{Available for download from \url{http://surveys.roe.ac.uk/ssa/}.} contains 1.9 billion of sources, with information about photometry, morphology and  quality of measurements. 
For our purposes we selected sources with a detection in both $B$ and $R$ bands (discarding the shallower $I$ plates). 
Unlike in \cite{2MPZ} and \cite{WISC15}, we use here both extended and point sources from the SCOS database. 
To maintain uniformity, SCOS objects were selected to limits of $B<21$ and $R<19.5$ in a pseudo-AB system (Peacock et al., in prep.).
As in WISE, we have also removed SCOS sources from the Galactic Plane ($|b|<10^\circ$), where blending and high Galactic extinction make the SCOS photometry unreliable.

\begin{figure}
  \centering
  \includegraphics[width=0.73\textwidth]{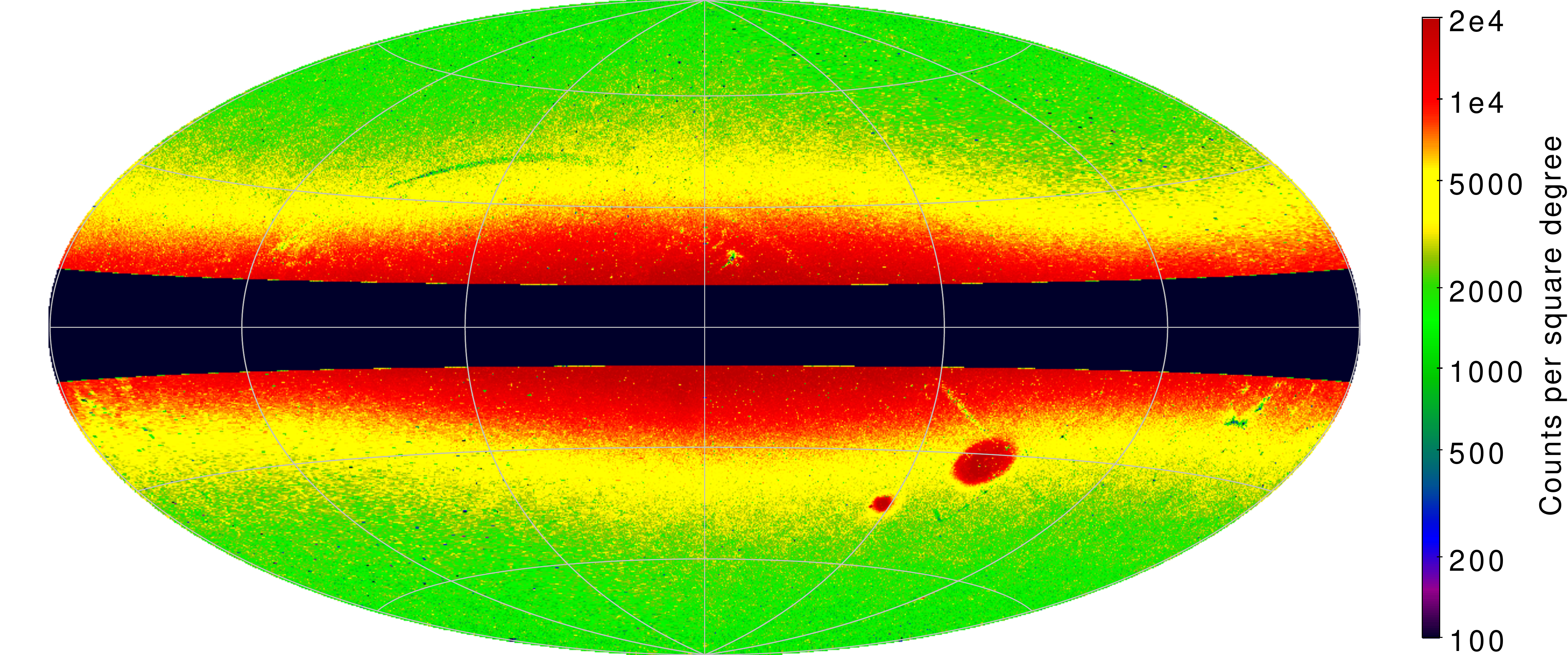}
  \caption{WISE$\times$SCOS catalog to which the classification is applied. Illustrated are 170 million sources in an all-sky Aitoff projection in Galactic coordinates.}
  \label{fig:1}
\end{figure}

The two photometric catalogs were paired up using a matching radius of $2"$.  
The resulting flux-limited WISE$\times$SCOS sample at $|b|>10^\circ$ contains almost 170 million sources, and is illustrated in Fig.~\ref{fig:1}. 
For a more comprehensive description of the catalogs used to construct the main photometric dataset (WISE and SCOS), see \cite{2MPZ,WISC15}, and Bilicki et al.\ in this volume. 

\section{Method: Support Vector Machines}

The main concept of Support Vector Machines (SVM) is to  calculate decision planes between a set of objects having different class memberships, the planes being defined by quantities that describe the properties of each class provided in a training sample. 
SVM searches for the optimal separating hyperplane between
the different object classes by maximizing the margin
between the classes' closest points, and the objects are classified based on their  position relative to the separation boundary in the
$n$-dimensional parameter space.
In our analysis, to search for the hyperplane we use a Gaussian kernel function, and a soft-boundary SVM method called C-SVM, which employs two parameters: $C$, giving a trade-off between large margins of different object classes, and $\gamma$, that determines the topology of the decision surface. 
Both parameters need to be tuned based on the
training sample. 
We employed LIBSVM, 
integrated software for support vector classification, implemented in the R 
software environment\footnote{\url{http://www.csie.ntu.edu.tw/cjlin/~libsvm} and  \url{http://www.r-project.org}} 
for statistical computing. 

\subsection{Training sample}
The successful application of SVM depends on a well-chosen training sample. 
As the algorithm searches for patterns between different object classes, the method heavily relies on a well-defined, representative sample of sources with already known properties. 
To train the classifier for the WISE$\times$SCOS catalog, we decided to use information from the spectroscopic sample of the Sloan Digital Sky Survey (SDSS DR12, \citealp{SDSS.DR12}), cross-matched with our photometric data. 
Pairing up WISE$\times$SCOS with SDSS DR12 sources  within 1’' we obtained a training sample of over 1.3 million common objects. % (68\% of galaxies, 25\% of stars, and 7\% of quasars).
Due to the SDSS spectroscopic preselection, such a training sample suffers from the lack of galaxies at the faint end of our catalog ($W1>16$); we have thus used the oversampling technique to improve the final selection. 
For more detailed description of the oversampling procedure we refer the reader to \cite{malek13SVM}. 

Our full training set was composed of 25,000 galaxies, 25,000 quasars, and
25,000 stars, selected as described below.
Fig.~\ref{fig:2} (\textit{upper left panel}) shows the number counts of the final training sample as a function of $W1$. 
From the  remaining objects from the WISE$\times$SCOS$\times$SDSS catalog randomly selected  (without repetition) 75,000 objects (galaxies, quasars and stars) were used for independent tests of classifier performance.

\subsection{SVM training details}

To  separate more efficiently different groups of objects and improve the classification process, we have divided the training sample into five  apparent magnitude bins:  $W1< 13$, $13 \leq W1 < 14 $, $14 \leq W1 < 15$, $15\leq W1 < 16 $ and $16 \leq W1 < 17 $. 
This allowed us to define the best classifiers as a function of $W1$, and  fully exploit even small differences between galaxies, stars, and quasars in 5 different magnitude bins. 
Therefore, our final SVM classifier is a combination of 5 different ones, and depending on the object's $W1$ magnitude, a suitable classifier is used.  

For the SVM parameter space we chose five quantities:
$W1$ magnitude, $W1$-$W2$  $R$-$W1$,  and $B$-$R$ colors and a $W1$ differential aperture magnitude (the latter helping separate point-like from resolved sources). 
Each classifier was tuned to find the best $C$ and $\gamma$ parameters, based on 5,000 galaxies, 5,000 quasars, and 5,000 stars selected randomly from the WISE$\times$SCOS$\times$SDSS catalog  in the appropriate $W1$ range.

\section{Results}

To quantify the efficiency of the WISE$\times$SCOS classifier we counted true objects (TG: true galaxies; TQ: true quasars; TS: true stars) from the training sample, classified properly by SVM, as well as false sources (FG: false galaxies, being in fact quasars or stars, misclassified as galaxies; etc.\ for FQ and FS), and calculated the total accuracy of our classifier from the formula:
\begin{equation}\label{TotalAccuracy}
\mathrm{TA}=\frac{1}{N}\sum_{i=1}^{N}\frac{TG+TQ+TS}{TG+TQ+TS+FG+FQ+FS}, 
\end{equation}
where $N$=10 is the number of cross-validation iterations performed for all five classifiers independently.  

\begin{figure}
  \centering
     \includegraphics[width=0.342\textwidth,height=0.18\textheight]{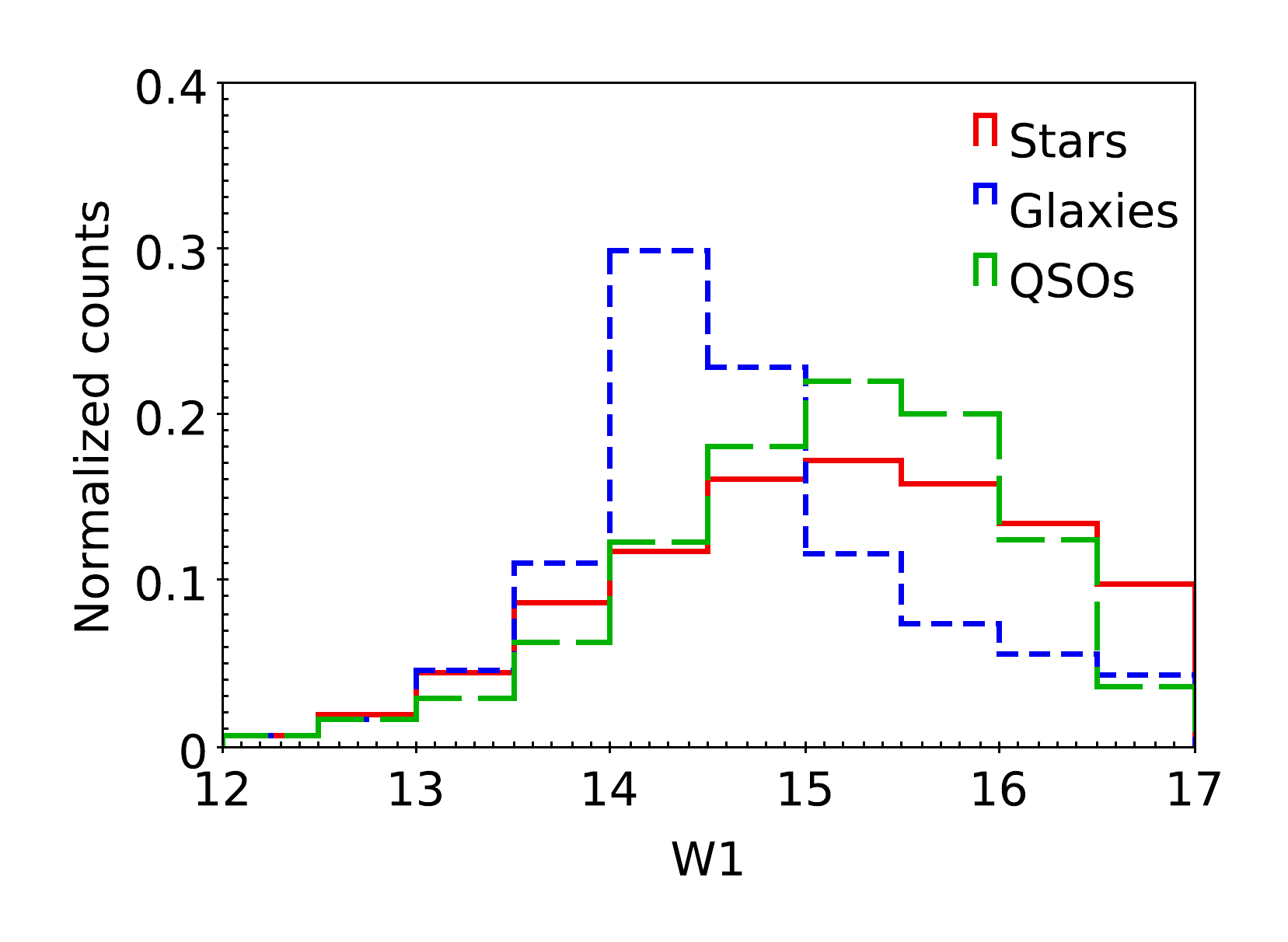}\includegraphics[width=0.342\textwidth,height=0.18\textheight]{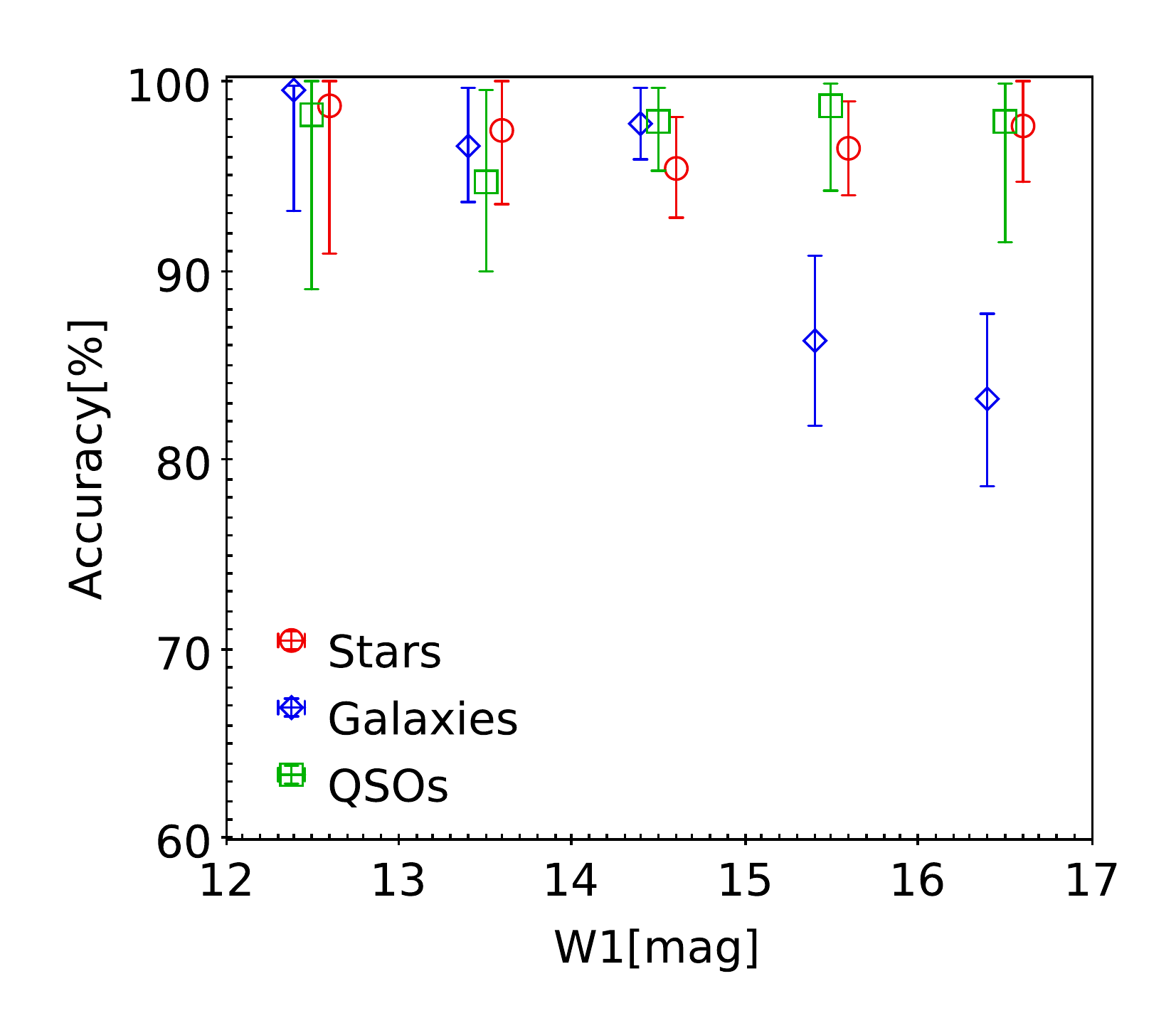}\includegraphics[width=0.342\textwidth,height=0.18\textheight]{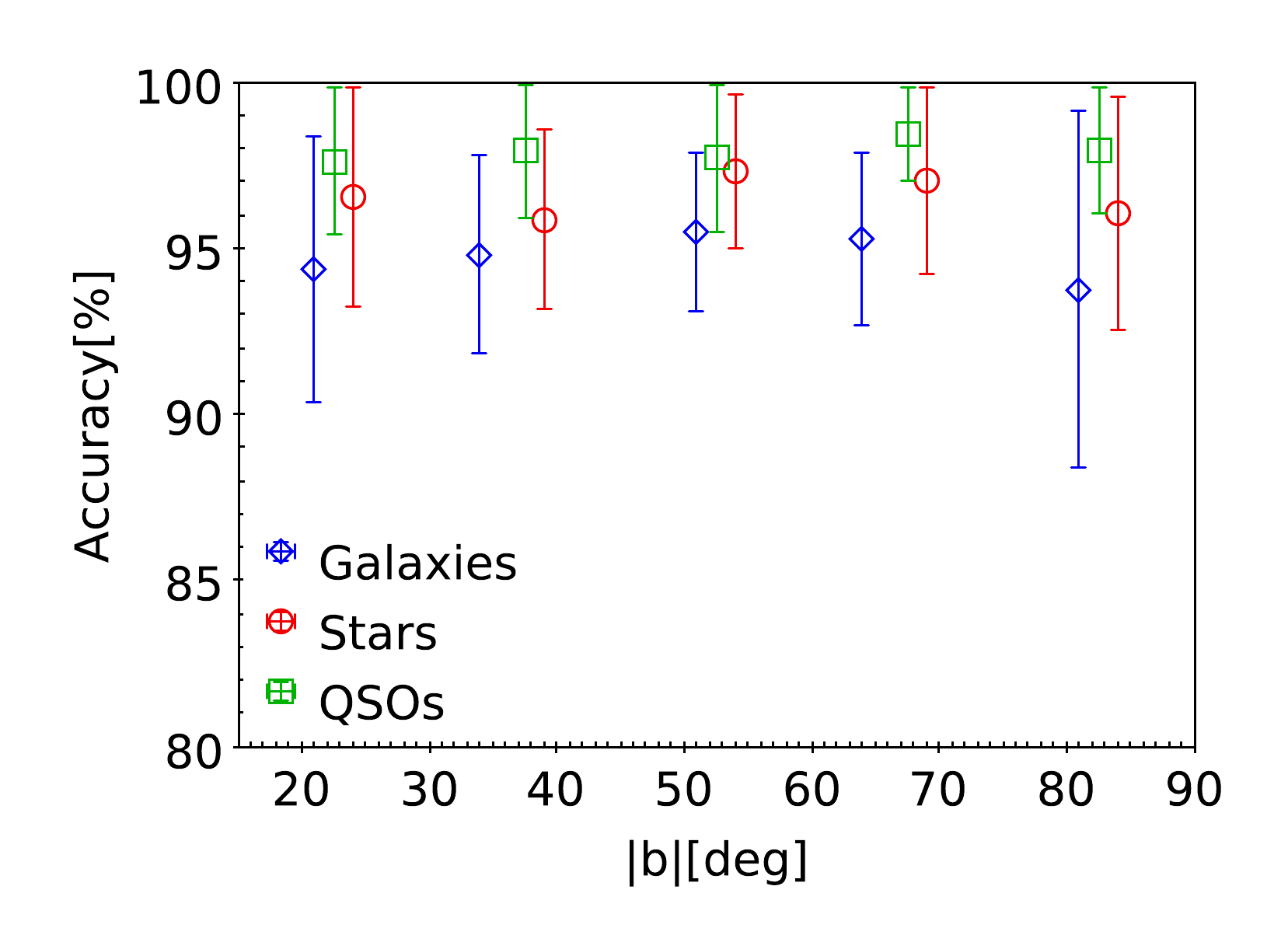}
     \includegraphics[width=0.342\textwidth,height=0.18\textheight]{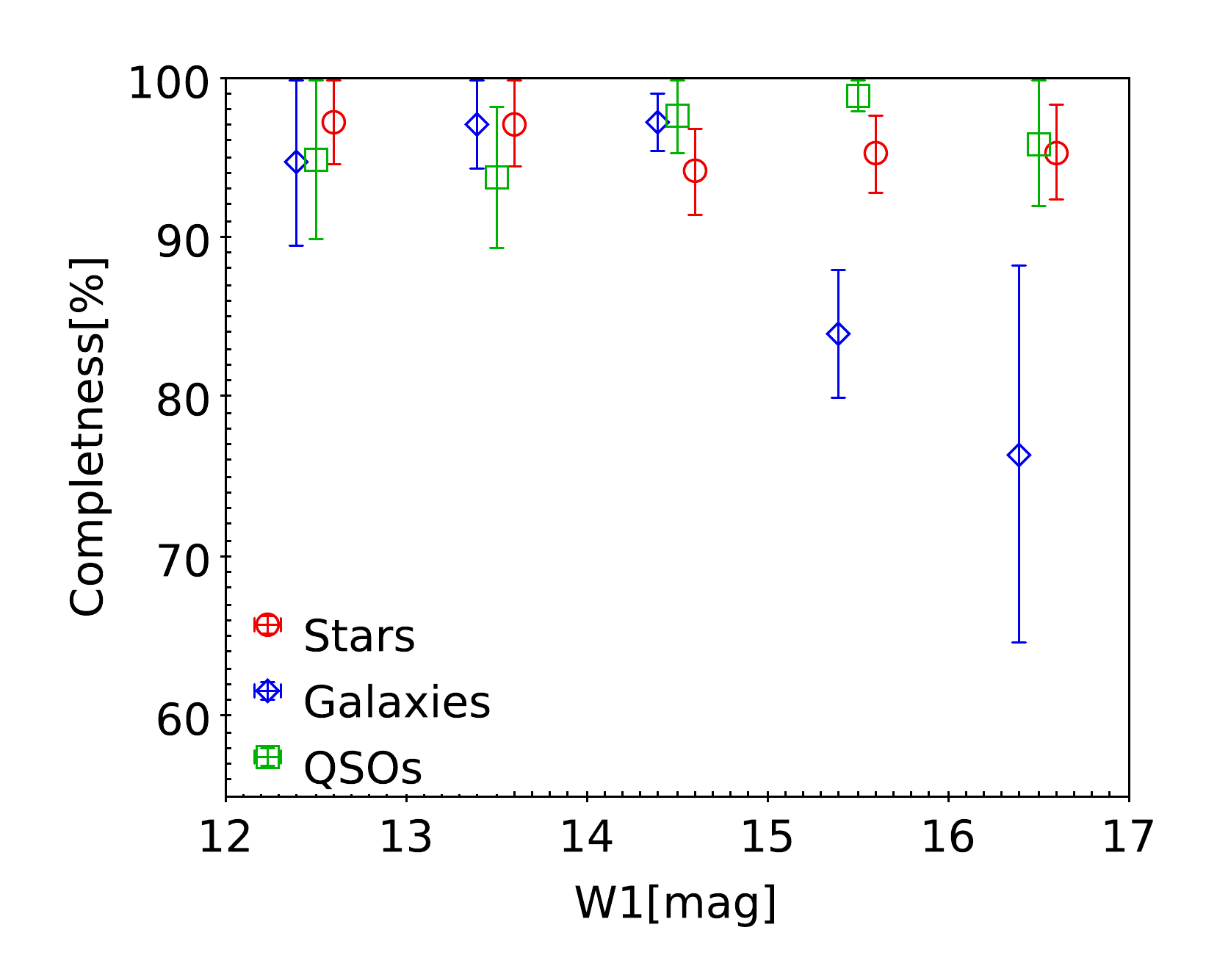}\includegraphics[width=0.342\textwidth,height=0.18\textheight]{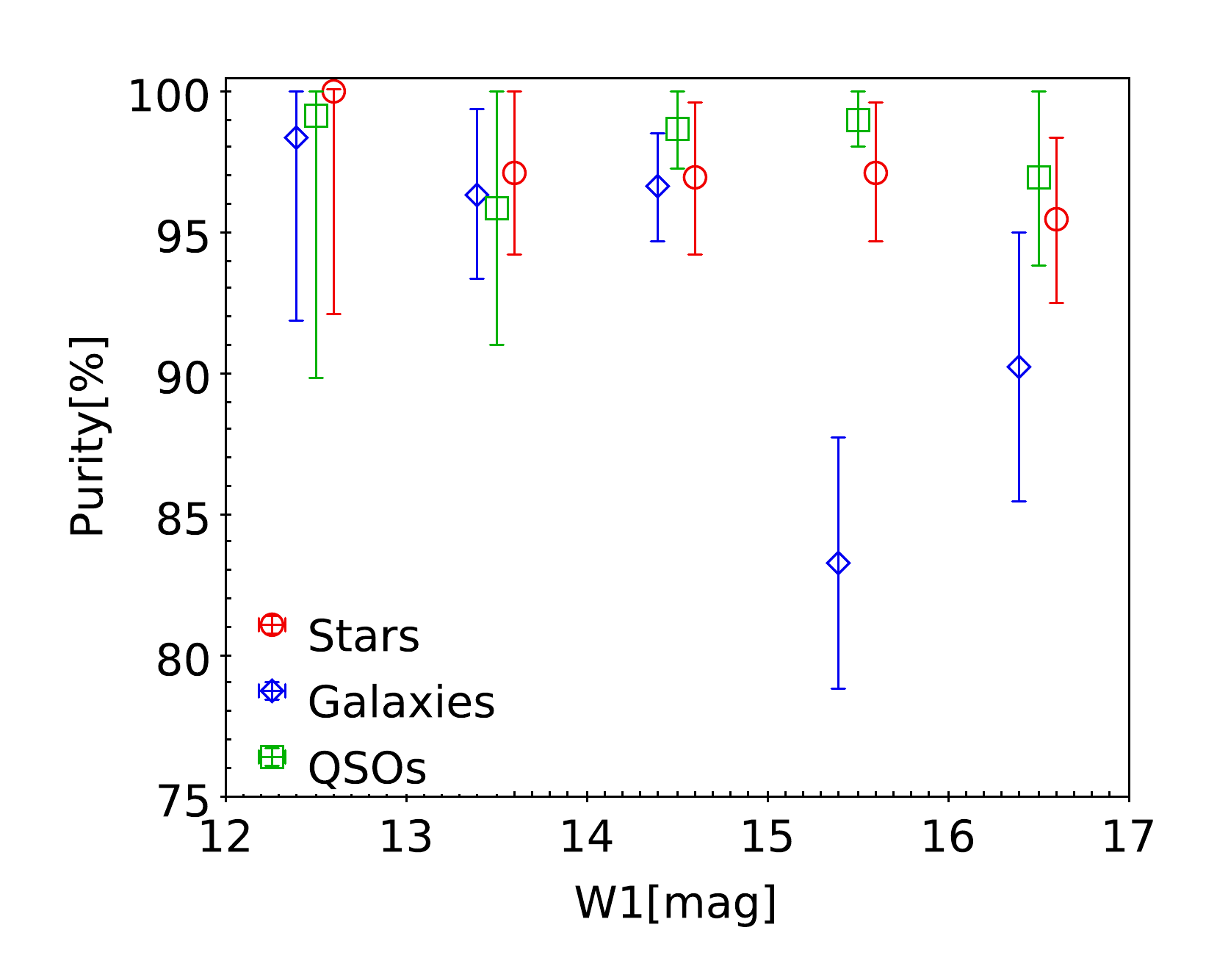}
     \caption{\textit{Top left:} Number counts as a function of $W1$ magnitudes in the WISE$\times$SCOS$\times$SDSS training sample, after oversampling of the galaxy population at the faint end. \textit{Top central and right:} Total accuracies for a self-check of the classifier, as a function of $W1$ magnitude and Galactic latitude $|b|$. \textit{Bottom:} Completeness (\textit{left}), and purity (\textit{right}), for the different source types, as a function of $W1$ magnitude.}
    \label{fig:2}
\end{figure}

We checked how the accuracy of the classifiers depends on the $W1$ magnitude, as well as on the Galactic latitude (Fig.~\ref{fig:2}, \textit{upper central and right panels}). 
We have found that in general the accuracies retain very high levels, of the order of 95\%, but there is significant deterioration in the efficiency for faint galaxies.  
The accuracy for galaxies with $W1 > 15.5$ mag is still at a very good level, more than 80\%, but the decline of the classifier's performance is visible. 
This is related to the fact that beyond $W1\gtrsim 15.5$ mag, the training set contains very few galaxies, and the oversampling procedure was not sufficient to compensate for this imbalance (Fig.~\ref{fig:2}, \textit{left panel}).
In most cases of incorrect classification, the true galaxies are misclassified as stars, and vice versa. 

To check the classifier's reliability for different ranges of the Galactic latitude, we calculated the accuracy in six $15^\circ$-wide bins in $|b|$.
For the latitude $|b|>15^\mathrm{o}$ the accuracy is always higher than 90\% for all the 3 classes. 
This shows that outside of the Galactic Plane our classifier does not depend on the latitude, and can be reliably used there. 
We have also examined the purity and completeness  of the training sample (see \citealt{soumagnac13} for relevant definitions), which are illustrated in the bottom panels of Fig.~\ref{fig:2}, and found generally very high levels of the two statistics, $>90\%$. The exception are galaxies fainter than $W1\sim 15$ mag, for which both completeness and purity decrease respectively to $\sim80\%$ and $\sim85\%$. This is again related to the lack of such sources in the training sample, which cannot be fully compensated for by the oversampling procedure.

\subsection{Final classification}

Having examined in detail the properties of our classifier, we applied it to the full WISE$\times$SCOS dataset preselected as described above. The final catalogs required some cleaning, such as removal of high-extinction regions ($EBV>0.5$ mag), and of the areas by the Galactic Plane and Bulge where blends dominate (cf.\ Fig.\ \ref{fig:1}). In addition, to maintain reliability, we took advantage of classification probabilities provided by SVM. 
We required the probability of each given class to be greater than 0.5, i.e.\ $p_{\rm{SVM}}(\rm{gal})>0.5$ for galaxies and similarly for stars and quasars.

The above criteria resulted in over 24 million galaxy and 3.4 million quasar candidates identified in the almost full-sky WISE$\times$SCOS sample, the remaining sources being either star candidates, or unclassified (the latter when for each type, its probability was $p(\rm{type})<0.5$).

We conclude that applying the SVM algorithm to WISE data cross-matched with SCOS, using additional source type information from the SDSS spectroscopic survey as a training set, can deliver uniform and reliable all-sky catalogs of galaxies, quasars and stars. 
One should however keep in mind that at present, the SCOS data do not provide photometry for point and extended sources calibrated in the same way. 
For that reason, in the forthcoming paper \citep{Krakowski16} we will focus on identifying galaxies using only the resolved SCOS sources cross-matched with WISE, as in \cite{WISC15}. 
In \cite{Krakowski16} we also provide more details of this work, such as various tests of the classifier, as well as the final catalog of WISE$\times$SCOS galaxy candidates identified in an automatized way.

\acknowledgements{This work was supported by the Polish National Science Center under contracts \# UMO-2012/07/D/ST9/02785 (KM, MB, AP, MK, AK \& AS), UMO-2013/09/D/ ST9/04030 (KM \& TK) and UMO-2015/16/S/ST9/00438 (AS). AP was partially supported by the Polish-Swiss Astro Project, co-financed by a grant from Switzerland, through the Swiss Contribution to the enlarged European Union. 
Special thanks to Mark Taylor for the TOPCAT
%\footnote{\url{http://www.star.bristol.ac.uk/$\sim$mbt/topcat/}}, 
and STILTS
%\footnote{\url{http://www.star.bristol.ac.uk/$\sim$mbt/stilts/}}  
 software.}

\bibliographystyle{ptapap}
\bibliography{KMalek_bib}

\end{document}